\documentstyle[11pt,newpasp,twoside,epsf]{article}
\def\edcomment#1{\iffalse\marginpar{\raggedright\sl#1\/}\else\relax\fi}
\reversemarginpar

\def\bout{B_{\rm out}}
\def\ceff{C_{\rm eff}}
\def\simlt{\mathrel{\hbox to 0pt{\lower 3.5pt\hbox{$\mathchar"218$}\hss}
      \raise 1.5pt\hbox{$\mathchar"13C$}}}
\def\simgt{\mathrel{\hbox to 0pt{\lower 3.5pt\hbox{$\mathchar"218$}\hss}
      \raise 1.5pt\hbox{$\mathchar"13E$}}}

\begin{document}

\title{The role of outflows and star formation efficiency in the evolution
of early-type cluster galaxies}
\author{Ignacio Ferreras \& Joseph Silk}
\affil{Nuclear \& Astrophysics Laboratory, Keble Road, Oxford OX1 3RH, U.K.}

\begin{abstract}
A phenomenological model for chemical enrichment in early-type galaxies
is presented, in which the process of star formation is reduced to a set
of four parameters: star formation efficiency ($\ceff$), fraction
of ejected gas in outflows ($\bout$), formation redshift ($z_F$) and
infall timescale ($\tau_f$). Out of these four parameters, only variations
of $\bout$ or $\ceff$ can account for the color-magnitude relation.
A range of outflows results in a metallicity sequence, whereas a
range of star formation efficiencies will yield a mixed age + metallicity
sequence. The age-metallicity degeneracy complicates the issue of
determining which mechanism contributes the most (i.e. outflows 
versus efficiency). However, the determination of the slope of the 
correlation between mass-to-light ratio and mass in clusters at moderate 
or high redshift will allow us to disentangle age and metallicity.
\end{abstract}


\section{Introduction}
One of the long-standing problems in astrophysics is the process of star
formation in galaxies. The standard scenario assumes stars to form 
from gas that falls in the potential wells of dark matter halos. 
Subsequent interacting or merging stages among galaxies might trigger
additional bursts of star formation. The complex nature of star formation
makes this problem rather an untractable one from an analytical point
of view, so that the best approach towards understanding the distribution
of stellar populations in galaxies requires a heavy use of rough 
approximations and all-too-often dangerous generalizations. 
It is the purpose of current phenomenological models describing the 
formation and evolution of the stellar component in galaxies to 
reveal the mechanisms which describe the wide range of galaxy colors 
and luminosities as well as their connection to morphology.
The current status of the determination of the ages of the stellar 
populations in galaxies is rather controversial due to the
degeneracy between age and metallicity (Worthey 1994). Observations of 
early-type galaxies by two different groups using similar techniques
targeting narrow spectral indices to infer a luminosity-weighted
age give contradictory results. While Trager et al. (2000) find
a large age spread in the sample of field and group early-type systems
of Gonz\'alez (1993), Kuntschner (2000) reports a large metallicity
spread in Fornax cluster ellipticals. So far, any observational
measurement of age is plagued by many degeneracies which render a direct
estimate uncertain. An alternative approach modelling the formation
and chemical enrichment of the stellar component of galaxies
is needed in order to reveal the actual scenario of galaxy formation.

\begin{figure}
 \epsfxsize=2.6in
\begin{center}
 \leavevmode
 \epsffile{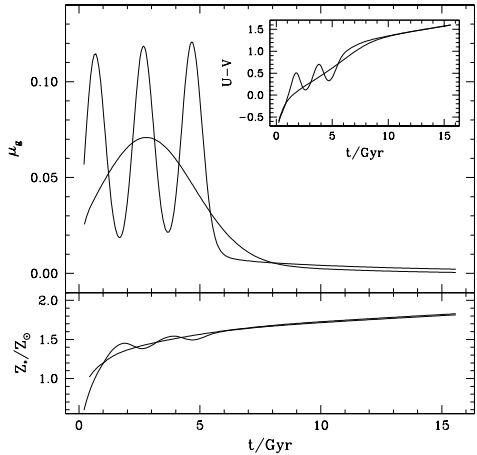}
\end{center}
\vskip -0.3truein
 \caption{Gas ({\sl top}) and metallicity ({\sl bottom}) evolution
of two models: one with a series of short bursts ({\sl thin line})
and another with a single, more extended burst. The inset shows
the evolution of $U-V$ color in the rest frame. 
}\label{f1}
\end{figure}

\section{Modelling chemical enrichment}
The model presented here describes the process of star formation
in early-type galaxies in terms of four parameters: star formation
efficiency ($\ceff$), ejected gas fraction in outflows ($\bout$),
formation redshift ($z_F$) and infall timescale ($\tau_f$).
The latter two parameters refer to the epoch ($t = t(z_F)$) 
at maximum and spread of a Gaussian profile for the
infalling gas, i.e.:
\begin{equation}
f(t) \propto e^{-\left(t-t(z_F)\right)^2/2\tau_f^2}
\end{equation}
This gas will be turned into stars according to a
linear Schmidt-type law, where the proportionality constant is
the star formation efficiency parameter. The model is described in
more detail in Ferreras \& Silk (2000a,b). This generic description 
allows us to include multi-burst scenarios in galaxies undergoing
several merging stages with enough gas to fuel star formation
at each merging event. Figure~1 shows a comparison between two star
formation histories: one with three equally strong starburst events 
and a second one as a simplification to the former using our
approach. The top panel shows the evolution of the gas mass --- which
is proportional to the star formation rate. The bottom panel
shows the evolution of the metallicity and the top inset traces
the evolution of rest frame $U-V$ color for both scenarios.
One can see that at times after the last bursting episode,
the evolution in both cases is roughly undistinguishable. Hence,
we conclude that our four-parameter model can account not only for a
standard ``monolithic'' scenario but also for multi-burst formation
histories. Furthermore, a Gaussian profile for infall avoids the
overproduction of low metallicity stars. In fact, a suitable choice 
of infall parameters ($\tau_f$,$z_F$) can reproduce the local 
metallicity distribution of stars (Rocha-Pinto \& Maciel 1996).

\begin{figure}
 \epsfxsize=3.5in
\begin{center}
 \leavevmode
 \epsffile{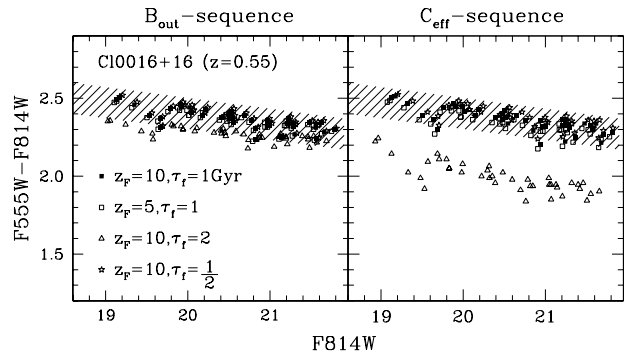}
\end{center}
\vskip-0.3truein
 \caption{Evolution of the color-magnitude relation observed in Coma
(Bower, Lucey \& Ellis (1992) projected to moderate redshift using both
an outflow-driven sequence ($B_{out}$, {\sl left}) and an efficiency-driven
sequence ($C_{eff}$, {\sl right}). The shaded area corresponds to the
observations of cluster Cl0016+16 ($z=0.55$, Ellis et al. 1997).
}\label{f2}
\end{figure}

For a given set of four parameters ($\bout$,$\ceff$,$z_F$,$\tau_f$) 
we can trace a star formation history and convolve it in age and 
metallicity with the simple stellar populations of Bruzual \& Charlot
(in preparation). Hereafter a closed cosmology ($\Omega_m=0.3$, 
$\Omega_\Lambda =0.7$, $H_0=60$ km s$^{-1}$ Mpc$^{-1}$),
and a hybrid Initial Mass Function between Scalo and Salpeter 
(Ferreras \& Silk 2000b) are used. Out of these four parameters, 
we find only $\bout$ and
$\ceff$ can generate the range of observed $U-V$ colors in nearby
early-type cluster galaxies (Bower, Lucey \& Ellis 1992). Hence, 
the luminosity sequence of these systems could be explained either
by a range of outflows ($\bout$-sequence), by a range of star formation
efficiencies ($\ceff$-sequence), or by some combination thereof.
Figure~2 shows the predicted color-magnitude relation (CMR) of 
Coma galaxies at the redshift of cluster Cl0016+16 
($z=0.55$, Ellis et al. 1997) assuming a $\bout$-sequence 
({\sl left}) or a $\ceff$-sequence ({\sl right}), and a range
of infall parameters ($z_F$,$\tau_f$). A sequence driven by
$\ceff$ results in an age spread for the stellar populations. This
causes the remarkable departure of the predictions from the
observed CMR (shaded area) for extended star formation histories
($z_F=10$, $\tau_f=2$ Gyr). However, because of the age-metallicity
degeneracy, we find that quite a large range of the parameters 
agree with the observations within error bars. Hence, we cannot use 
photometric measurements of moderate redshift clusters in order
to determine whether age ($\ceff$) or metallicity ($\bout$)
drive the CMR.

\section{$M/L$ ratio as tracer of age evolution}
One of the most age-sensitive observables is the mass-to-light
ratio. Hence, the predicted evolution of $M/L$ with lookback
time should be different for sequences driven by age or
by metallicity. We consider the evolution with redshift 
of the slope of the correlation
between $M/L$ in rest frame $B$-band and {\sl stellar} mass.
This slope change is parametrized by $\eta_B$ defined as follows:
\begin{equation}
\eta_B(z) \equiv \left.\frac{\Delta\log M/L_B}{\Delta\log M}\right|_{z} -
	\left.\frac{\Delta\log M/L_B}{\Delta\log M}\right|_{z=0}
\end{equation}
For a $\bout$-sequence (driven by outflows), the range of luminosities is 
related by a spread in metallicities. As we evolve the cluster to 
higher redshifts, the mass-to-light ratio will decrease uniformly
across the luminosity sequence because of lookback time, and there
will also be a relative change of $M/L$ among early-type galaxies 
caused by its very weak metallicity dependence,
which makes the decrease in mass-to-light ratio slighly larger in
galaxies with a higher metallicity, thereby flattening the slope
of $M/L$ vs $M$ (i.e. $\eta_B\simlt 0$). On the other hand,
a $\ceff$-sequence (driven by efficiency) will introduce a significant
age spread which varies with galaxy mass, so that $M/L$ at the 
fainter end (which has a lower efficiency and thus a larger age 
spread) will decrease more than the bright end, steepening the
slope (i.e. $\eta_B > 0$, see figure~5 in Ferreras \& Silk 2000b).

This behavior makes the study of the evolution of $M/L$ with
redshift a suitable candidate to infer the star formation
history of early-type cluster galaxies. Unfortunately, this
observable still poses a long string of uncertainties which prevent it
from establishing a clearcut way of breaking the degeneracy between
age and metallicity: mass-to-light ratios require time-consuming 
spectral observations in order to measure velocity dispersions,
and can only be achieved with 10m class telescopes for clusters
at moderate and high redshifts. Furthermore, the measured $M/L$ ratios
(inferred from observations of velocity dispersions, 
surface brightnesses and galaxy sizes) rely on a set of assumptions 
about the structure of the galaxy. Any correlation between galaxy 
structure and mass or luminosity will add systematic errors which are 
hard to estimate. However, alternative age-dependent observables 
such as Balmer spectral indices are also plagued by model-dependent 
uncertainties. Despite all these caveats, the study of the evolution 
of $M/L$ with lookback time is still one of the best methods to determine 
the stellar demography in galaxies.

\end{document}